\documentclass{article}

\usepackage{listings}
\usepackage{amsmath}
\usepackage{graphicx}
\usepackage{PRIMEarxiv}
\usepackage{environ}

\usepackage[utf8]{inputenc} 
\usepackage[T1]{fontenc}    
\usepackage{hyperref}       
\usepackage{url}            
\usepackage{booktabs}       
\usepackage{amsfonts}       
\usepackage{nicefrac}       
\usepackage{microtype}      
\usepackage{lipsum}
\usepackage{fancyhdr}       
\usepackage{graphicx}       
\graphicspath{{media/}}     

\title{MCMChaos: Improvising Rap Music with MCMC Methods and Chaos Theory
}

\author{
  Robert Kimelman \\
  Johns Hopkins University \\
  \texttt{rkimelm1@jh.edu} \\
}

\begin{document}
\maketitle

\begin{abstract}
A novel freestyle rap software, MCMChaos 0.0.1, based on rap music transcriptions created in previous research is presented. The software has three different versions, each making use of different mathematical simulation methods: collapsed gibbs sampler and lorenz attractor simulation. As far as we know, these simulation methods have never been used in rap music generation before. The software implements Python Text-to-Speech processing (pyttxs) to convert text wrangled from the MCFlow corpus into English speech. In each version, values simulated from each respective mathematical model alter the rate of speech, volume, and (in the multiple voice case) the voice of the text-to-speech engine on a line-by-line basis. The user of the software is presented with a real-time graphical user interface (GUI) which instantaneously changes the initial values read into the mathematical simulation methods. Future research might attempt to allow for more user control and autonomy.
\end{abstract}


\section{Introduction}
There has been a great deal of previous research on rap music
lyric generation \cite{malmi, potash}; but, as pointed out in other research \cite{xue}, this is not fully representative of the intricacies of rap music, as it contains complex rhyme schemes, intonation patterns, and other dynamically changing characteristics. In addition, comparatively less research has been done on rap music generation that includes real-time manipulation of parameters and involves an end user. Nonetheless, there have been very good efforts so for in this domain \cite{Wu, Torres}; we hope to take a new approach to this using unique mathematical simulation and rhythmic-making techniques. Our software is also open-source, so that it can be used for other research and potentially educational purposes. Markov Chain Monte Carlo (MCMC) methods are widely used for computational rap music analysis and generation \cite{addanki, conklin}, as well as several other statistical and machine learning methods for music generation as a whole \cite{kalingeri, mao, conklin}.

$\>$We propose a rap music generation software based on transcriptions in
MCFlow: A Digital Corpus of Rap Transcriptions \cite{conditschultz} called MCMChaos 0.0.1. The text versions of these transcriptions were obtained using a function written in the R programming language which is also integrated in the latest version of humdrumR: a user-friendly software for computational music analysis \cite{ConditSchultz2019}. What makes MCMChaos unique in the field of rap music generation is that it allows for an interactive experience that can be used for perceptual and computational studies of rap music as well as for STEAM educational purposes. Students can learn about mathematical simulation, programming, and music while having fun and developing their own independent creations. The code is also open-source, so students can tweak the code to their liking to experiment and observe how it works. We invite any collaborators and students to improve or build upon the code.
\section{Text Processing}
\label{sec:headings}

\qquad We obtained the corpus of rap music transcriptions from the following
\href{https://github.com/Computational-Cognitive-Musicology-Lab/MCFlow}{GitHub repository} \cite{conditschultz}. The transcriptions were originally in syllable format, so we wrote a function to convert them into text format. The code for this transformation is rather simple. The function takes in a boolean argument nullTokens which represents whether or not the user would like to output the corresponding text format with tokens representing where the additional syllables would be for a given word. The code for when nullTokens = TRUE is left out because we used nullTokens = FALSE for this software. The function begins by checking that the character vector read into the text function is in proper format:
\begin{center}
\begin{lstlisting}[language=R]
text <- function(data, nullTokens = TRUE){
  print(silbeFormat(data))
  # check that the char vector is in correct syllable format
  if(nullTokens == FALSE){
    # if the user does not want null tokens to replace instances of syllables 
    # occurring after the first syllable of a multi-syllable word
    data <- toString(data[,1])
    # transform character vector to string
    data <- stringr::str_replace_all(data, "-, -", "")
    # remove all instances of -, -, which represents a space between two syllables 
    # which when combined form a word
    data <- stringr::str_replace_all(data, ",", "")
    # remove all instances of , which occur after every word except the last one
    data <- as.list(strsplit(data, '\\s+')[[1]])
    # get all of the words as a list
    transpose1 <- t(data)
    transpose2 <- t(transpose1)
    data <- as.character(transpose2)
    # transform the data further to get desired character vector
  }
    if(nullTokens == TRUE)...
}
\end{lstlisting}
\end{center}

A mapping from the original format to text format looks like this, with both nullTokens = TRUE and nullTokens = FALSE shown for illustrative purposes: \\
  
\begin{center}
\begin{tabular}{ c c c c }
 original & & nullTokens = FALSE & nullTokens = TRUE \\
 Now &  &  & Now \\ 
 let &  &  & let\\  
 me &  &  Now & me\\
 wel- &  &  let & welcome\\
 -come &  &  me & .\\
 e- & $\rightarrow$ & welcome & everybody \\
 -very- &  &  everybody & .\\
 -bo- &  &  to & .\\
 -dy &  &  the & .\\
 to &  &  wild& to\\
 the &  &  wild & the\\
 wild &  &  west & wild\\
 wild &  &  & wild\\
 west &  & & west\\
 
\end{tabular}
\end{center}

\section{Collapsed Gibbs Sampler Implementation}
We chose to use Collapsed Gibbs Sampler specifically for its usefulness in simulating correlated data from a target distribution. In actual rap music, the performer should rap such that the characteristics of each line are somehow related; otherwise, the music would likely not sound rhythmic and pleasurable to listen to.

The target distribution and sampling method is as follows: \\
\begin{center}
    Simulate 
        $\begin{bmatrix} 
        x \\
        y \\
        z 
        \end{bmatrix} \sim N\Bigg(\begin{bmatrix}
        50 \\
        45 \\
        1 \\
        \end{bmatrix},
        \begin{bmatrix}
        1 & 0 & \rho \\
        0 & \rho & 0 \\
        \rho & 0 & 1 \\
        \end{bmatrix}
        \Bigg)$. Then we can estimate the variance as the determinant of the covariance matrix:
        $1(1(\rho)-0^2) + 0 + \rho(\rho^2)$.\\
        . \\
        A Gibbs Sampler might try to make use of the following distributions : \\
        $X\mid Y,Z \sim N(E[X\mid Y,Z], \sqrt{\rho + \rho^3})$  \\
        $Y\mid X,Z \sim N(E[Y\mid X,Z], \sqrt{\rho + \rho^3})$ \\
        $Z\mid X,Y \sim N(E[Z\mid X,Y], \sqrt{\rho + \rho^3})$ \\
        
        Collapsed Gibbs Sampling will instead sample from $x\mid y$ and $y\mid x$, and then we can infer $E[Z\mid X, Y]$ based on $E[x\mid y]$ and $E[y \mid x]$ and obtain a fairly accurate sample. \\
        
        We know that $f(x,y,z)$ can be written as $f(z\mid x, y)f(x,y)$. So, as stated, we can first sample from $x\mid y$ and $y\mid x$ and proceed to sample from $z\mid x,y$.\\
        Since we know $E[x\mid y] = \rho y$ and $E[y\mid x] = \rho x$, then we can estimate $E[z\mid x,y] = \rho x + \rho y$. \\
        Therefore, our Collapsed Gibbs Sampling Method makes use of the following distributions: \\
        
        $X\mid Y,Z \sim N(E[X\mid Y], \sqrt{\rho + \rho^3})$  \\
        $Y\mid X,Z \sim N(E[Y\mid X], \sqrt{\rho + \rho^3})$ \\
        $Z\mid X,Y \sim N(E[Z\mid X,Y], \sqrt{\rho + \rho^3})$ \\
        $=$\\
        $X\mid Y,Z \sim N(\rho*y, \sqrt{\rho + \rho^3})$  \\
        $Y\mid X,Z \sim N(\rho*x, \sqrt{\rho + \rho^3})$ \\
        $Z\mid X,Y \sim N(\rho*x + \rho*z, \sqrt{\rho + \rho^3})$ \\
        
\end{center}
Through trial and error, we were able to implement a Collapsed Gibbs 
Sampler version of the software which converges to the target distribution while printing values that fall within a specified range. We tweaked the sampling method in such a way because the voice type, volume, and rate of speech only take on a particular range of values (voice: $|$-48$|$ to 48, volume: $|$-1$|$ to 1, and rate of speech is only intelligible up to a certain value which differs depending on the voice type, but often $\approx$ 1 is the lower limit and $\approx$ 500 is the upper limit, with values between 1-50 being optimal depending on the voice. To test that our sampling method was converging to the target distribution, we simulated 10,000 samples with the given initial values of $x = 50$, $y = 45$, and $z = 1$. Our sampling method did indeed converge to the target normal distribution, as can be seen in the histograms below.

\begin{center}
\includegraphics[scale=0.28]{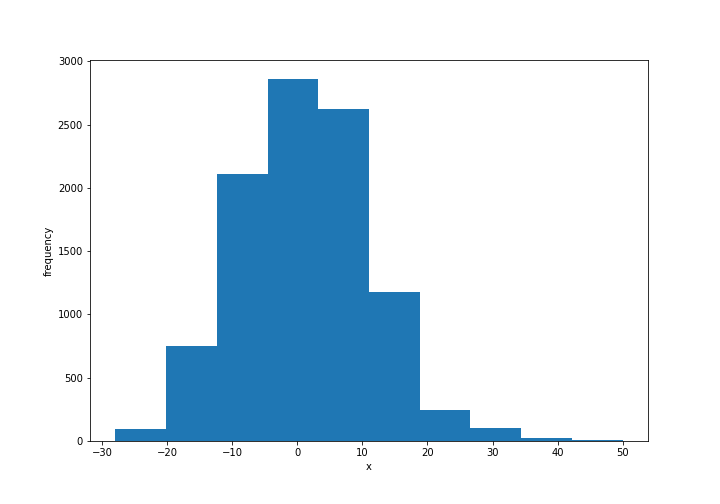}
\includegraphics[scale=0.28]{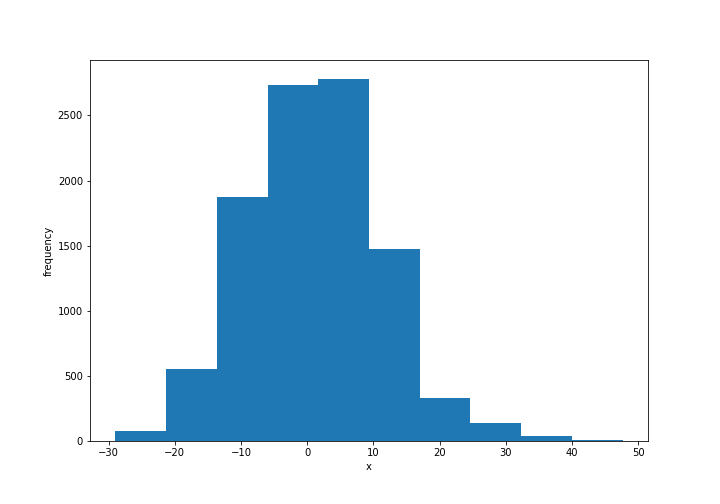}
\includegraphics[scale=0.28]{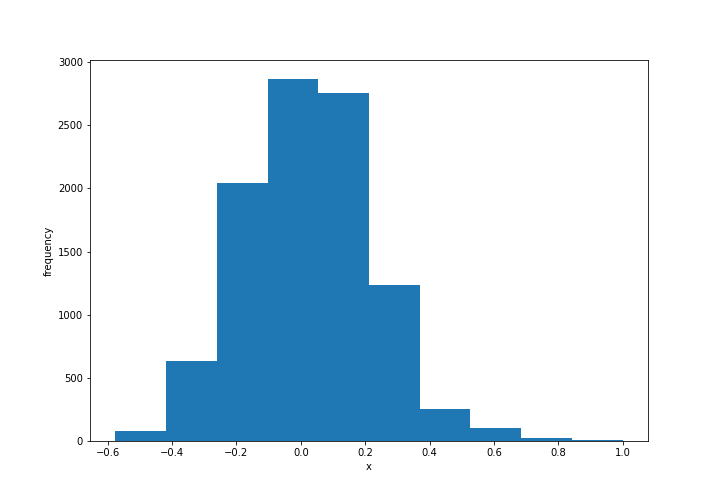}
\end{center}

To further verify our results, we did another 100 samples, and here we show trace plots of the paths that the simulation took for each value. These trace plots illustrate that our simulations are indeed converging.

\begin{center}
\includegraphics[scale=0.28]{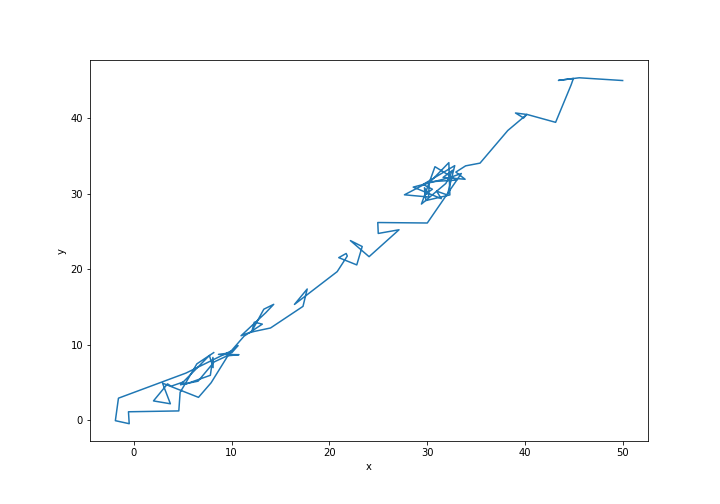}
\includegraphics[scale=0.28]{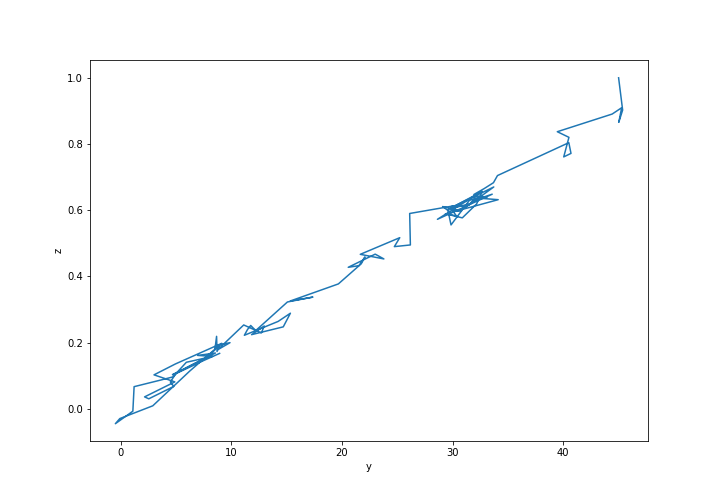}
\end{center}

The code for the simulation is as follows: 

\begin{lstlisting}[language=Python]
def collapsedgibbs3NormV(n, rho, x, y, z):
    mat = mat=np.zeros((n,3))
    mat[0, ] = np.array([x,y,z])
    for i in range(1,n):
        x = np.random.normal(rho * y, math.sqrt(rho+rho**3), 1)
        y = np.random.normal(rho * x, math.sqrt(rho+rho**3), 1)
        z = np.random.normal(rho * x + rho * y, math.sqrt(rho+rho**3), 1)/100
        mat[i, ] = [x,y,z]
    return(mat)
\end{lstlisting}

We divided z by 100 to ensure that the values will usually be $<$ 1, but this does not affect convergence to the normal distribution as shown in the plots above.

\subsection{Interface}

The text-to-speech function is as follows:
\begin{lstlisting}[language=Python]
def text_to_speech(numberOfLines = 1, startingPoint = 1, rho = 0.99, x = 50, y = 45, 
z = 0):
    engine = pyttsx3.init()
    engine.setProperty("volume", z)
    engine.setProperty("voice", voices[y].id)
    engine.setProperty("rate", x)
    if(z !=0):
        sample = collapsedgibbs3NormV(numberOfLines, rho, x, y, z)
        print(sample)
        for i in range(startingPoint, startingPoint + len(sample)):
            engine = pyttsx3.init()
            rows[i][0] = rows[i][0][1:]
            str1 = ''.join(rows[i])
            if(abs(sample[i-startingPoint,0]) >= 20 and 
            abs(sample[i-startingPoint,0]) <= 1000 and 
            abs(sample[i-startingPoint,0]) != 0):
                engine.setProperty("rate", abs(sample[i-startingPoint,0]))
            else:
                engine.setProperty("rate", 50)
            engine.setProperty("voice", voices[7].id)
            if(abs(sample[i-startingPoint,2]) <= 1):
                engine.setProperty("volume", sample[i-startingPoint,2])

            else:
                engine.setProperty("volume", 0.5)
            engine.say(str1)
            engine.runAndWait(); 
            engine.stop()
\end{lstlisting}
The code for multiple voices looks the same with the exception of the portion of the code responsible for setting the voice. In the multiple voice version, it is:
\begin{lstlisting}[language=Python]
if(int(sample[i-startingPoint,1]) > 47):
    engine.setProperty("voice", voices[17].id)
else:
    engine.setProperty("voice", voices[int(sample[i-startingPoint,1])].id)
\end{lstlisting}
The GUI looks like the following:
\begin{center}
\includegraphics[scale=0.6]{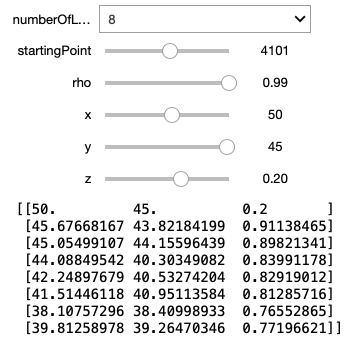}
\end{center}

The user can first alter the number of lines they'd like to be performed (max. = 10), the starting point of the rap in the MCFlow dataset \cite{conditschultz}, the correlation between the values ($\rho$), and the initial x (rate of speech), y (voice), and z (volume) values read into the collapsed gibbs sampler simulation. Once the value for z is changed to a value other than 0, then the software will proceed to sample and utter the lines desired by the user. 

\section{Lorenz Attractor Implementation}
\NewEnviron{myequation}{%
    \begin{equation}
    \scalebox{2}{$\BODY$}
    \end{equation}
    }
The lorenz system can be modeled by the following differential equations:
\begin{center}
    \begin{myequation}%
    \frac{dx}{dt} = \sigma(y-x); \frac{dy}{dt} = x(\rho-z)-y; \frac{dz}{dt} = xy-\beta z %
    \end{myequation}

\end{center}

Implemented in Python, this is:
\begin{lstlisting}[language=Python]
def chaosTheoryXYZ(n,time, sigma, rho, beta, x, y, z):
    pointFinal = []
    for i in range(0,n):
        dx = (sigma * (y - x))*time
        dy = (x * (rho - z) - y)*time
        dz = (x * y - beta * z)*time
        x = x + dx
        x = get_digit(x,0)*100
        y = y + dy
        z = z + dz
        z = get_digit(z,0)/10
        
        point = [x, y, z]
        pointFinal += [point]
    return(pointFinal)
\end{lstlisting}

This is a bit of an adaptation of the original lorenz attractor system and takes a simple approach to solving the differential equations \cite{youtubevideo}. Since this simulation eventually yields very large numbers, we've chosen to modify it such that only the first digit of x and y are used and then multiplied by 100 and $\frac{1}{10}$ respectively in order to obtain values that fall within the permissible range \cite{stackoverflow}.

\subsection{Interface}

The interface looks the same as that with the collapsed gibbs sampler implementation, but with more values that can be altered and a slightly different output:

\begin{center}
\includegraphics[scale=0.6]{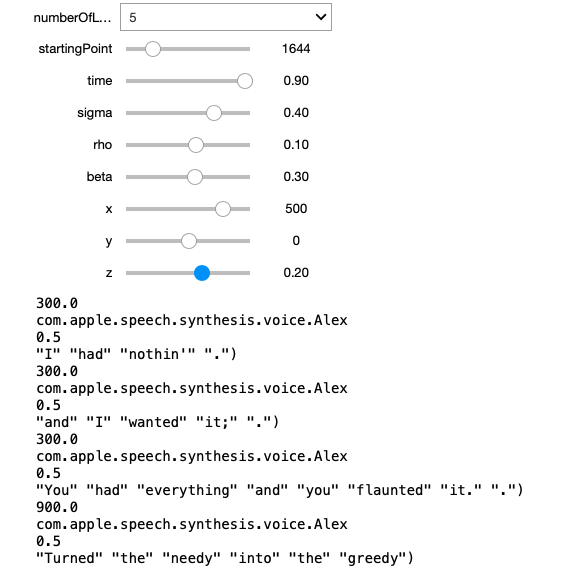}
\end{center}

\section{Own Voice Implementation}

In addition, we have also included an implementation that allows the user to rap in their own voice with the help of a previous implementation \cite{siri}. This implementation utilizes the CMU Pronouncing Dictionary and sounds made by the user of the software (current implementation has Kimelman's voice). The user can type in any word or phrase they'd like to hear. The software can also be modified to work with the MCFlow dataset \cite{conditschultz} and potentially in tandem with either mathematical simulation, which will be included in a later version of the software.

\section{Simplistic Yet (More) Rhythmic and Realistic Implementation}

This implementation does not make use of either mathematical simulation method, but instead chooses a random number between 170-200 as the rate of speech for voice number 7 for particular line segments broken into either $\frac{1}{4}$, $\frac{1}{3}$, or $\frac{1}{2}$ size. As an example, the Python code for the $\frac{1}{2}$ size version is as follows:
\begin{lstlisting}[language=Python]
def text_to_speech(numberOfLines = 1, startingPoint = 1, rho = 0.99, x = 50, y = 45, z = 0):
    # 1/2 version
    engine = pyttsx3.init()
    engine.setProperty("volume", z)
    engine.setProperty("voice", voices[7].id)
    engine.setProperty("rate", x)
    if(z !=0):
        sample = collapsedgibbs3NormV(numberOfLines, rho, x, y, z)
        for i in range(startingPoint, startingPoint + len(sample)):
            engine = pyttsx3.init()
            rows[i][0] = rows[i][0][1:]
            str1 = ''.join(rows[i])
            str1 = str1.replace("(", "")
            str1 = re.split('\s+', str1)
            engine.setProperty("rate", randrange(170,200))
            rate = engine.getProperty("rate")
            if(rate > 185):
                str2 = str(str1)
                str2 = str2.replace(",","")
                engine.say(str2)
                engine.runAndWait();
                engine.stop()
                print(str2)
            else:
                str2 = str(str1[0:int(len(str1)/2)])
                str2 = str2.replace(",", "")
                engine.say(str2)
                engine.setProperty("rate", (5/4)*rate)
                engine.setProperty("volume", z+0.15)
                str3 = str(str1[(int(len(str1)/2)):len(str1)])
                str3 = str3.replace(",","")
                engine.say(str3)
                engine.runAndWait();
                engine.stop()
                print(str3)
\end{lstlisting}

This version was made with the intention to achieve a more "realistic" (i.e., rhythmic) version that sounds more like a human rapper. In the original mathematical simulation methods, the changing values alter the utterances line-by-line, but since rapping typically includes an oscillating and changing voice within lines, we decided to include this simpler implementation which sounds a bit more rhythmic.

\section{RoboRapRhythm Implementation}

In this implementation, an automatic lyrics-to-time aligner with millisecond precision was used \cite{Gupta} to program a text-to-speech algorithm such that each line was said at their respective starting points. The first song used is called "Welcome to the Party" by rapper Pop Smoke.

\section{Musical Properties}

This software was written with the intent to embody specific musical characteristics, some specifically in relation to musical improvisation. First, given the importance of the presence of silence in music, the software accounts for that and allows for values of 0 in the simulations. In addition, given the nature of the step-wise process in which computers read code, there is a constant pause between each rap line as the computer takes a new sample from one of the mathematical simulations. This is a vital, well-known characteristic of rap music, in which the performer takes a slight pause in between each line for emphasis. This pause is one of the primary distinctions between rapping and speaking.

Furthermore, in improvisation, there is an inherent feature of randomness that one hopes will culminate in order. Both the lorenz attractor and collapsed gibbs sampler implementations are chaotic and random; yet the sampling distribution of the collapsed gibbs sampler follows a consistent normal distribution, with more predictable, but still nonetheless random, outputs and the lorenz attractor 3D graph results in an oscillatory plane that looks like a butterfly (hence, the butterfly effect).

Interestingly, our simplest implementation is the most nuanced in terms of its musical properties. One of the main goals of this project was to generate a voice that can be reliably recognized as a rapping voice. As we were prototyping the software, we realized that it could be made more musical if there were variations of the rapping voice within lines. If this were not present in rap music, one would likely find it less entertaining and attention-grabbing.

\section{Obstacles and Future Directions}

This is a first prototype; thus, there are many improvements that can be made. The voices used are not as rhythmic as we'd like them to be. To create a software that raps such that it is nearly indistinguishable from a human rapper, we predict that this will likely take a novel text-to-rap algorithm that does not make use of previous text-to-speech processors, which we hope to include in a later version.

In addition, it takes some getting used to the simulations for the user to understand how it works and get random yet efficient samples that result in a desired behavior (for example, chaotic).

Furthermore, while making the software, it was difficult to find mathematical simulations that both obey both mathematical laws and fit into the allowed range of values for the Python Text-to-Speech parameters. There were many working prototypes that obeyed one set of laws but not the other and vice versa, and some of course that did not obey either set.

We also hope to integrate human-computer interaction and perceptual studies with this software, such as making surveys that measure a user's level of expressivity, enjoyment, and control of the software, as well as whether or not audio clips of this software are actually perceived as rap music (or poetic).

\section{Acknowledgments}

The author would like to thank Alexander Cohen and Nathaniel Condit-Schultz for their guidance and advice on this paper.

\bibliographystyle{unsrt}

\end{document}